\documentclass[aip,amsmath,amssymb,reprint]{revtex4-1}

\usepackage{soul}
\usepackage[normalem]{ulem}
\usepackage{threeparttable}
\usepackage{graphicx,color,upgreek}
\usepackage{mathtools}
\usepackage{amsmath}
\usepackage{graphicx,color,natbib}
\usepackage{placeins}
\usepackage[dvipsnames]{xcolor}

\usepackage{enumitem} 

\setlist{nolistsep}

\newcommand{\pecl}{\operatorname{\mathit{P\kern-.08em e}}}

\begin{document}

\title{Active Brownian Particles in Random and Porous Environments}
\date{\today}

\author{Fergus Moore}
\affiliation{Bristol Centre for Functional Nanomaterials, University of Bristol, Bristol BS8 1FD, United Kingdom}
%\affiliation{H.\ H.\ Wills Physics Laboratory, University of Bristol, Bristol BS8 1TL, United Kingdom}
\affiliation{H.H. Wills Physics Laboratory, Tyndall Ave., Bristol, BS8 1TL, UK}

\author{John Russo}
\email{\text{john.russo@uniroma1.it}}
\affiliation{Department of Physics, Sapienza University of Rome, P.le Aldo Moro 5, 00185 Rome, Italy}
\affiliation{School of Mathematics, University of Bristol, Bristol BS8 1UG, United Kingdom}

\author{Tanniemola B. Liverpool}
\affiliation{School of Mathematics, University of Bristol, Bristol BS8 1UG, United Kingdom}

\author{C. Patrick Royall}
\email{\text{paddy.royall@espci.psl.eu}}
\address{Gulliver UMR CNRS 7083, ESPCI Paris, Universit\' e PSL, 75005 Paris, France.}
\affiliation{School of Chemistry, Cantock's Close, University of Bristol, BS8 1TS, UK}
\affiliation{H.H. Wills Physics Laboratory, Tyndall Ave., Bristol, BS8 1TL, UK}

\begin{abstract}
The transport of active particles may occur in complex environments, in which it emerges from the interplay between the mobility of the active components and the quenched disorder of the environment. Here we explore structural and dynamical properties of Active Brownian Particles (ABPs) in random environments composed of fixed obstacles in three dimensions. We consider different arrangements of the obstacles. In particular, we consider two particular situations corresponding to experimentally realizable settings.
Firstly, we model pinning particles in (non--overlapping) random positions and secondly in a percolating gel structure, and provide an extensive characterization of the structure and dynamics of ABPs in these complex environments. We find that the confinement increases the heterogeneity of the dynamics, with new populations of absorbed and localized particles appearing close to the obstacles. This heterogeneity has a profound impact on the motility induced phase separation (MIPS) exhibited by the particles at high activity, ranging from nucleation and growth in random disorder to a complex phase separation in porous environments.
\end{abstract}

\maketitle

\section{Introduction}
\label{sectionIntroduction}
Active matter concerns systems comprised of individual bodies undergoing motion via self-propulsion \cite{ramaswamy2017}. This description encompasses a plethora of systems over a wide range of lengthscales from bacteria \cite{zhang2010,lopez2015,jepson2013}, biological microswimmers \cite{elgeti2015}, schools of fish \cite{yang2022}, bird flocks \cite{cavagna2014}, to human crowds \cite{koyama2020}.  While these systems can all be classified as active matter, accurate models must be tailored to the specifics of each system and its environment. One class of much simpler model systems which nevertheless captures key elements of the behavior of more complex systems is active colloids  \cite{bechinger2016,buttinoni2013,bricard2013,mauleonamieva2020}.

Yet if we are to apply such model systems in a biological context, it is essential that we study the dynamics of active particles in environments that are relevant to their real-world counterparts. For biological active matter on mesoscopic lengthscales this means environments like porous soils \cite{gannon1991} and organic tissues \cite{isermann2017}. Environments such as these have several qualities in common, they are often crowded, random and irregular. This of course has an impact on the transport or displacement of the active bodies inside these spaces \cite{bechinger2016}, for example in the (biological) case of bacteria, their run--and--tumble dynamics can be drastically altered \cite{bhattacharjee2019,irani2022}.

In equilibrium systems, the dynamics of fluids in dense and complex environments has long been an area of interest. For example, the inclusion of specific structures into dense fluids has proven significant for progress towards understanding the glass transition and in liquids \cite{alcoutlabi2005,alba-simionesco2006}. Furthermore, the addition of randomly pinned particles within a dense ensemble is known to greatly slow the dynamics of such systems, providing access to rare states \cite{russo2010,karmakar2013,kim2011,ozawa2015}. A special kind of localisation has been explored in glass-forming systems \cite{biroli2008,cammarota2012}, and which may be realized using colloidal systems \cite{gokhale2016,gokhale2014,williams2018}. Here \textit{pinning}, i.e. immobilizing a fraction of the particles, provides access to the so--called ideal glass, a putative amorphous state of very low configurational entropy whose diverging timescales render it otherwise inaccessible to experiment or computer simulation \cite{berthier2011}.

For active systems, in the simplest case of a confining wall, self-propelling spheres will accumulate at a wall as a consequence of the timescale of its persistent motion, even in the absence of hydrodynamic interactions \cite{vanteeffelen2008,elgeti2013}. For many-body systems in two-dimensions, the influence of disordered landscapes on to the dynamics of active systems has been shown to manifest in clogging and localisation transitions \cite{chepizhko2019,reichhardt2017,reichhardt2018}, subdiffusion over long timescales \cite{chepizhko2013,zeitz2017,morin2017}, destruction of flocking clusters \cite{morin2017b}, suppression of Motility--Induced Phase Separation (MIPS) and the prevention of uniform wetting at boundaries \cite{dor2021}. Furthermore, manipulation of complex environments has been shown to provide a degree of control over the transport of active matter in the form of sorting \cite{volpe2011}, and the intriguing phenomenon of topotaxis (control over net flow directions by controlling the topology of the environment \cite{borba2020,schakenraad2020}). Active Brownian particles (ABPs) exhibit rich phase behavior such as MIPS \cite{wysocki2014,stenhammar2014}, and the formation of active crystals \cite{omar2021,moore2021} and fundamental properties such as pressure and the equation of state differ drastically from what might be expected from equilibrium systems \cite{mallory2014,bialke2015,solon2015}.

However, the question of how active Brownian spheres couple to a complex surrounding environment remains unanswered. Recently, there is been interest in experiments with mesoscale active matter in 3d. In one study a random heterogeneous environment was found to impose strong inhibitions on active transport of bacteria \cite{bhattacharjee2019}, and in another study the impact of dimensionality was made clear, with the dominance of three-dimensional structure in the presence of an anisotropic potential \cite{sakai2020}. This latter example is a well--controlled 3d colloidal model system which provides the inspiration for our work because it is possible to confine such systems using pinning \cite{gokhale2014,hermes2011} or allowing a subset of particles to undergo gelation \cite{varrato2012}. Insight into the transport of active matter in 3d complex environments could provide a major step towards control of such systems and aid progress towards applications such as drug delivery.

In the present article, we perform three-dimensional molecular dynamics simulations of active Brownian particles in complex heterogeneous environments. To model such environments, we choose two example structures which, as noted above, may be realized in experiment: a random homogeneous array of pinned particles, providing an extension of disordered random obstacle studies to 3D; and a continuous, percolating porous network (a gel), simulating the complex environments typical of active matter under confinement. Furthermore, we will investigate the structural and dynamical properties of ABPs within these structures, with a focus on phase separation and how this varies from the MIPS observed in bulk suspensions.

This article is organized as follows: in section \ref{sectionModel} we outline the computational methods used to study these systems, in section \ref{sectionResults} we discuss the results of the simulations, and finally in section \ref{sectionConclusion} we will summarize and conclude our findings and discuss future work in this area.

\section{Model and methods}
\label{sectionModel}

\begin{figure*}
\centering
\includegraphics[width=\linewidth]{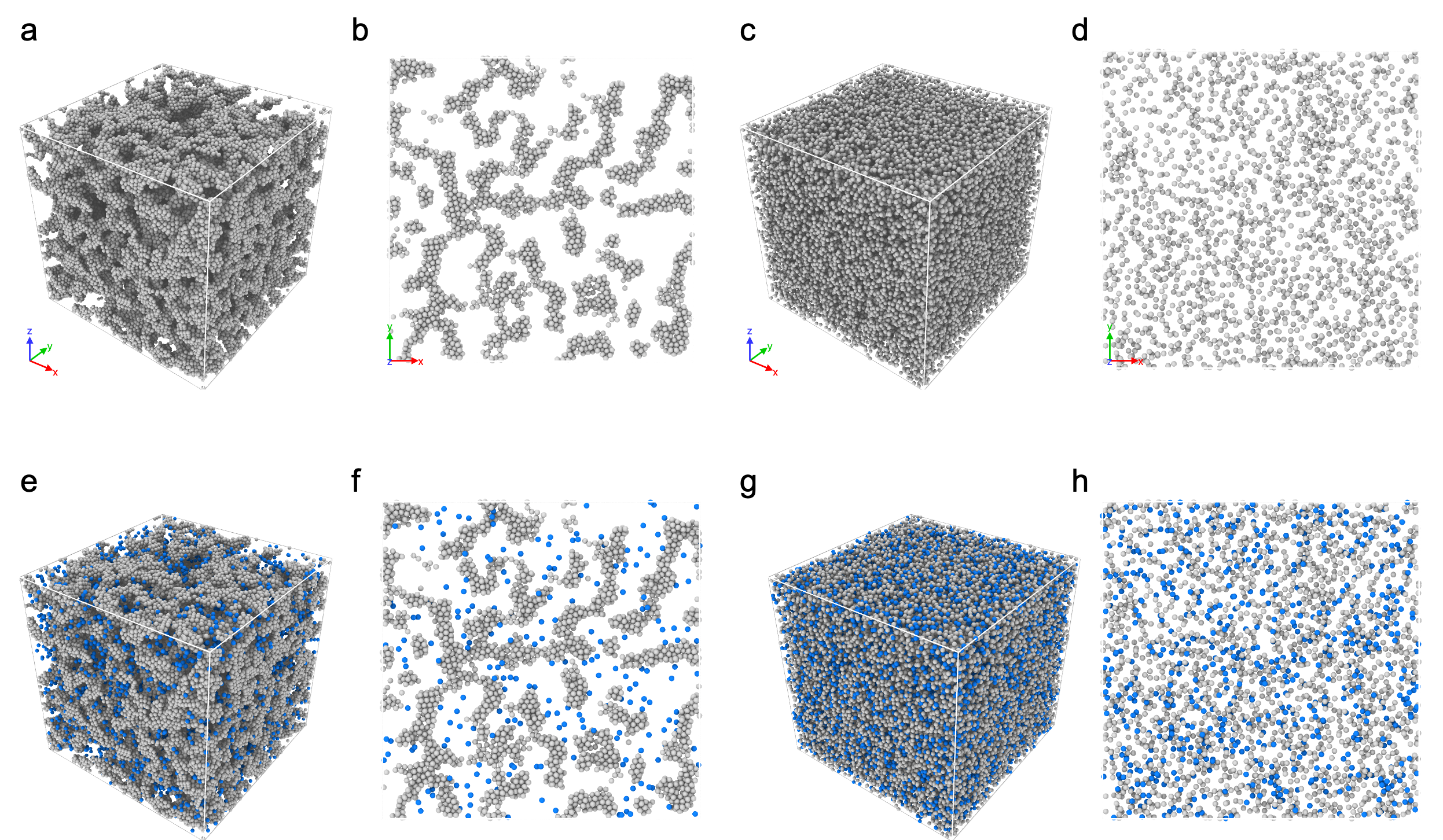}
\caption{Preparation of complex environments for ABPs.
(a) Percolating gel network at number density $\rho= 0.38$. 
(b) Cross-section through the gel network of depth 2$\sigma$.  (c) A collection of randomly pinned obstacles at $\rho =0.31$. (d) Cross-section through the random obstacles of depth 2$\sigma$. (e-f) Gel network (grey) filled with active particles (blue) to a total density $\rho=0.42$. (g-h) Random obstacles filled with active particles to a total number density $\rho=0.42$.}
\label{figObstacles}
\end{figure*}

\subsection{Active Particles}

We model active colloids as active Brownian particles, which propel with a constant velocity $V_{0}$ along their individual direction vectors $\mathbf{e}$, which in turn are subject to rotational diffusion.  We implement this model through molecular dynamics simulations using a customized version of the open source LAMMPS package \cite{plimpton1995,moore2021}, which integrates the following equations of motion:

\begin{equation}
\dot{\mathbf{r}}=V_{0} \mathbf{e}+\beta D_{T}\mathbf{F} +\sqrt{2D_{T}}\mathbf{\eta}
\label{eqEom1}
\end{equation}

\begin{equation}
\dot{\mathbf{e}}=\sqrt{2D_{R}}\mathbf{\xi} \times \mathbf{e}
\label{eqEom2}
\end{equation}

\noindent
Here $\dot{\mathbf{r}}$ is the particle velocity, $V_{0}$ is the magnitude of the constant active velocity, and $\mathbf{F}$ is the inter-particle force. The thermal fluctuations promoting translational diffusion are included in the Gaussian white-noise term
$\mathbf{\eta}$, where $\langle\mathbf{\eta}\rangle=0$, and $D_{T}$ is the translational diffusion coefficient. Thermal noise driving rotational diffusion of the direction vector $\mathbf{e}$ is represented by an independent Gaussian noise term $\mathbf{\xi}$,  where $\langle\mathbf{\xi}\rangle=0$, and $D_{R}$ is the rotational diffusion coefficient. The two diffusion coefficients are related via $D_{T} = D_{R}\sigma^{2}/3$, where $\sigma$ is the particle diameter. Time is scaled in units of the characteristic rotational diffusion time $\tau_R = 1 / (2D_R)$ \cite{wysocki2014}.  The active particles are modelled as being similar to hard spheres and to achieve this we include a Weeks-Chander-Andersen (WCA) inter-particle potential, which takes the form:

\begin{equation}
\beta u(r_{ij}) = \begin{cases}
4 \beta \varepsilon\left[\left(\frac{\sigma}{r_{ij}}\right)^{12} - \left(\frac{\sigma}{r_{ij}}\right)^{6}\right] + \beta \varepsilon & r_{ij} \leq 2^{\frac{1}{6}}\sigma \\%&\text{se $\omega\in A$}\\
0 & r_{ij} > 2^{\frac{1}{6}}\sigma   
\end{cases}
\label{eqWCA}
\end{equation}

\noindent
where $\varepsilon=5$ is the interaction strength $r_{ij}$ is the inter-particle distance, and $\beta=1/k_BT$ the thermal energy. Since we use the WCA interaction, 
it is hard to define a volume fraction. Methods that determine an effective diameter such as Barker--Henderson \cite{barker1967}, may not hold outside of equilibrium systems. Therefore, as in ref. \cite{martin-roca2021}, we use the total density $\rho = N / V$, where $N$ is the number of particles and $V$ is the volume of the system.

We use the P\'{e}clet number to refer to the relative strength of the activity in the system, which we define as: $\pecl = V_{0}/\sigma D_{R}$. Throughout this work we keep $D_R$ constant via $D_T =1$, and vary $\pecl$ by changing the propulsion velocity $V_0$.

Simulations are performed with periodic boundary conditions. The majority of the work is carried out in a cubic box of dimension length $L = 55\sigma$, with a total number of $N = 144000$ particles. In some cases, there was a need to sample from many state points and for these a smaller system was used where $L = 27.5\sigma$ and $N$ ranges from 18000 to 24000. Analysis at constant density is always conducted at $\rho = 0.87$. This state point is chosen such that it lies below the freezing line in the bulk.

\subsection{Complex Environments}
The complex environments relevant to microscopic biological systems are often irregular and random in nature. To investigate the dynamical properties of ABPs under these conditions, one must prepare obstacle geometries that satisfy these requirements. Here we consider two primary structures; porous gel networks and randomly pinned particles. As noted above, these may be realized in experiments. In addition to these, we include simulations studying the bulk dynamics of ABPs as a reference, these bulk simulations use the approach outlined in the previous section.

In the following section, we describe how the two complex environments are created and characterized. A schematic depicting these two environments is displayed in Fig. \ref{figObstacles}, featuring 3d renderings of each environment type along with a cross-sectional slice. The porous gel network (Fig. \ref{figObstacles}a-b), is a heterogeneous system, comprised of two distinct meso--phases which percolate through the entire simulation box comprised of the particle--rich phase and a particle--poor phase in which for our parameters no particles are found. The random environment (Fig. \ref{figObstacles}c-d), is comprised of randomly pinned particles that create number of discrete obstacles dispersed throughout the system.

\textit{Preparation of a porous network---}
In this work the porous network is modeled as a colloidal gel, specifically a colloid-polymer mixture \cite{poon2002,royall2021}. We do this because we seek to connect our work to experiments where such a network could be realized  \cite{royall2021,sakai2020}. Therefore, the preparation protocol in our simulations is as close to one which might be experimentally realizable as possible. For suitable parameters, colloids with such an attractive interaction begin to phase separate via spinodal decomposition which is then arrested, leaving a bicontinuous network, ie a gel \cite{zaccarelli2007,royall2021}. The (polymer-induced) attraction between the colloidal particles that would be used in an experiment is here modeled with the Morse potential \cite{royall2008,taffs2010jpcm},  
\begin{equation}                                                                                                                                                                                                                                                                                                                                                                                                                                                                                                                                                                                                                                                                                                                                                                                                                                                                                                                                                                                                                                                                                                                                                                                                                                                                                                                                                                                                                                                                                                                                                                                                                                                                                                                                                                                                                                                                                                                                                                                                                                                                                                                                                                                                                                                                                                                                                                                                                                                                                                                                                                                                                                                                                                                                                                                                                                                                                                                                                                                                                                                                                                                                                                                                                                                                                                                                                                                                                                                                                                                                                                                                                                                                                                                                                                                                                                                                                                                                                                                                                                                                                                                                                                                                                                                                                                                                                                                                                                                                                                                                                                                                                                                                                                                                                                                                                                                                                                                                                                                                                                                                                                                                                                                                                                                                                                                                                                                                                                                                                                                                                                                                                                                                                                                                                                                                                                                                                                                                                                                                                                                                                                                                                                                                                                                                                                                                                                                                                                                                                                                                                                                                                                                                                                                                                                                                                                                                                                                                                                                                                                                                                                                                                                                                                                                                                                                                                                                                                                                                                                                                                                                                                                                                                                                                                                                                                                                                                                                                                                                                                                                                                                                                                                                                                                                                                                                                                                                                                                                                                                                                                                                                                                                                                                                                                                                                                                                                                                                                                                                                                                                                                                                                                                                                                                                                                                                                                                                                                                                                                                                                                                                                                                                                                                                                                                                                                                                                                                                                                                                                                                                                                                                                                                                                                                                                                                                                                                                                                                                                                                                                                                                                                                                                                                                                                                                                                                                                                                                                                                                                                                                                                                                                                                                                                                                                                                                                                                                                                                                                                                                                                                                                                                                                                                                                                                                                                                                                                                                                                                                                                                                                                                                                                                                                                                                                                                                                                                                                                                                                                                                                                                                                                                                                                                                                                                                                                                                                                                                                                                                                                                                                                                                                                                                                                                                                                                                                                                                                                                                                                                                                                                                                                                                                                                                                                                                                                                                                                                                                                                                                                                                                                                                                                                                                                                                                                                                                                                                                                                                                                                                                                                                                                                                                                                                                                                                                                                                                                                                                                                                                                                                                                                                                                                                                                                                                                                                                                                                                                                                                                             \beta u\left(r_{ij}\right)=\beta \varepsilon \exp \left[a_{0}\left(\sigma-r_{ij}\right)\right]\left(\exp \left[a_{0}\left(\sigma-r_{ij}\right)\right]-2\right)
\label{eqMorse}
\end{equation}

\noindent
where $a_0=33$ is a range parameter.

To create the gel structures in simulation, we begin with particles in a simple cubic crystal at the desired number density, and then evolve this system according to Brownian dynamics (Eq. (\ref{eqEom1}), for $V_0 = 0$), with the particles interacting via the Morse potential Eq. (\ref{eqMorse}). This system is then evolved for $5 \times 10^{7}$ integration steps, which is equivalent to 1200$\tau_B$ after which the system is frozen and no further movement is allowed. Here $\tau_B$ is the Brownian time $\tau_B=(\sigma /2)^2 / 6D_T$. An example of the resulting gel is shown in Fig. \ref{figObstacles}(a-b).

With the gel in place, the positions of the free particles are initialized via the Lubachevsky--Stillinger algorithm \cite{lubachevsky1990}. This method comprises the following steps: First, initial particle positions are randomly assigned. Then, the particles are slowly grown and displaced from an initial diameter $\sigma_\mathrm{in}=0.1$, to the desired diameter $\sigma=1$, such that these particles experience minimal overlaps with themselves and with the frozen gel particles. Additionally, to guard against the presence of any small but sufficient remaining particle overlaps, a pre-run simulation with a soft potential is performed:

\begin{equation}
u(r_{ij})=A\left[1+\cos \left(\frac{\pi r_{ij}}{r_{c}}\right)\right] \quad r_{ij}<r_{c}
\label{eqsoft}
\end{equation}

\noindent
where $r_c = 2^\frac{1}{6}$ is the potential cut-off, and the constant A is ramped from 0 to 100 over $1.2 \tau_R$ without activity. Following this the system is equilibrated again without activity, for particles following the equations of motion outlined in Eq. (\ref{eqEom1}) and Eq. (\ref{eqEom2}), and the WCA inter-particle potential Eq. (\ref{eqWCA}). A low density example of this system is shown in Fig. \ref{figObstacles}(e-f).

\begin{figure*}
\centering
\includegraphics[width=0.8\linewidth]{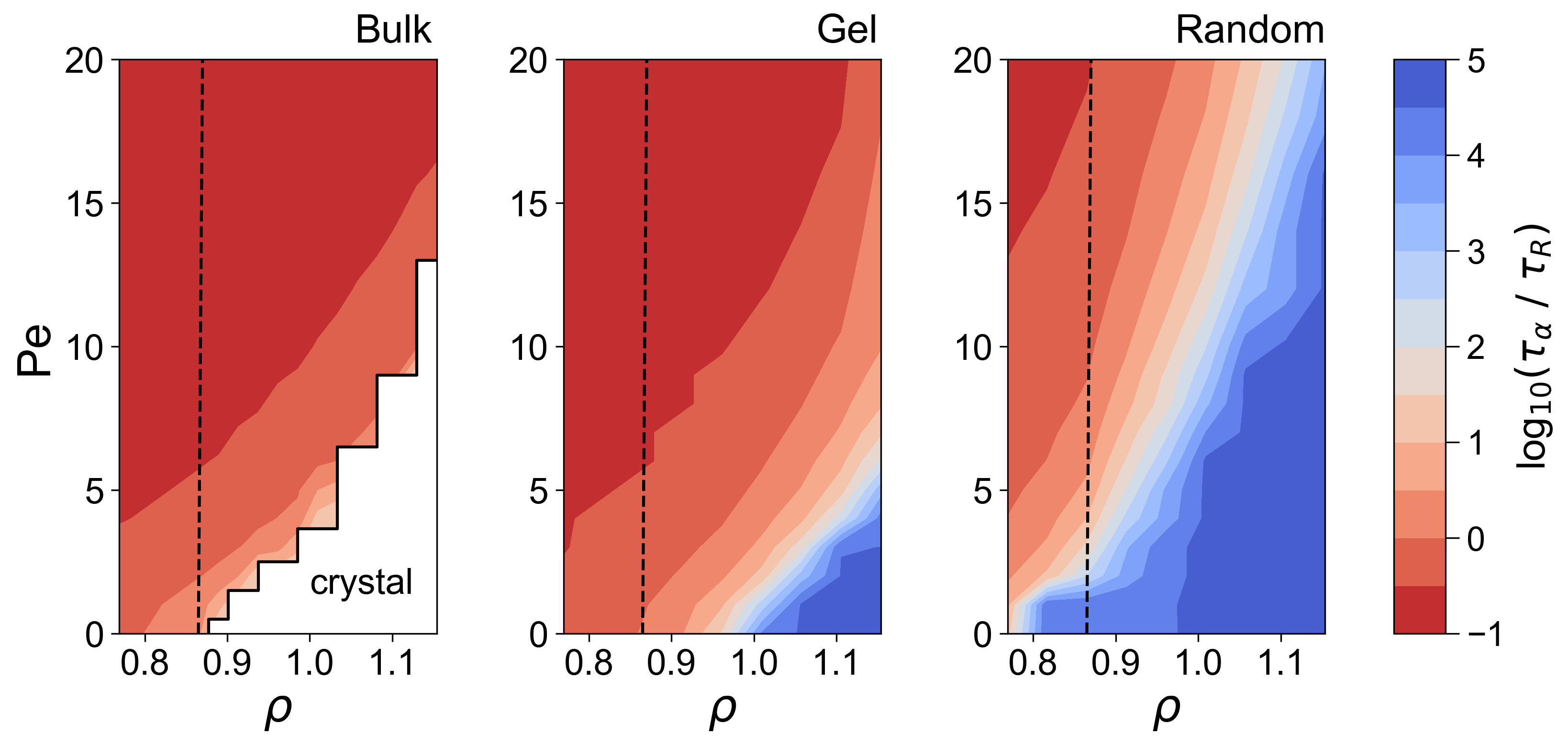}
\caption{Structural relaxation time ($\tau_{\alpha}$) as a function of number density  $\rho$ and P\'{e}clet number $\pecl$.  Here we consider the bulk, gel network and random pinning systems from left to right.
Colours indicate the common logarithm of $\tau_{\alpha}$. Crystalline states are marked in white. The number density $\rho = 0.87$ is emphasized with a black dotted line. Due to the large amount of sampling required, this data was measured using the smaller simulation box size.} 
\label{figTauAlpha}
\end{figure*}

\begin{figure}
\centering
\includegraphics[width=0.8\linewidth]{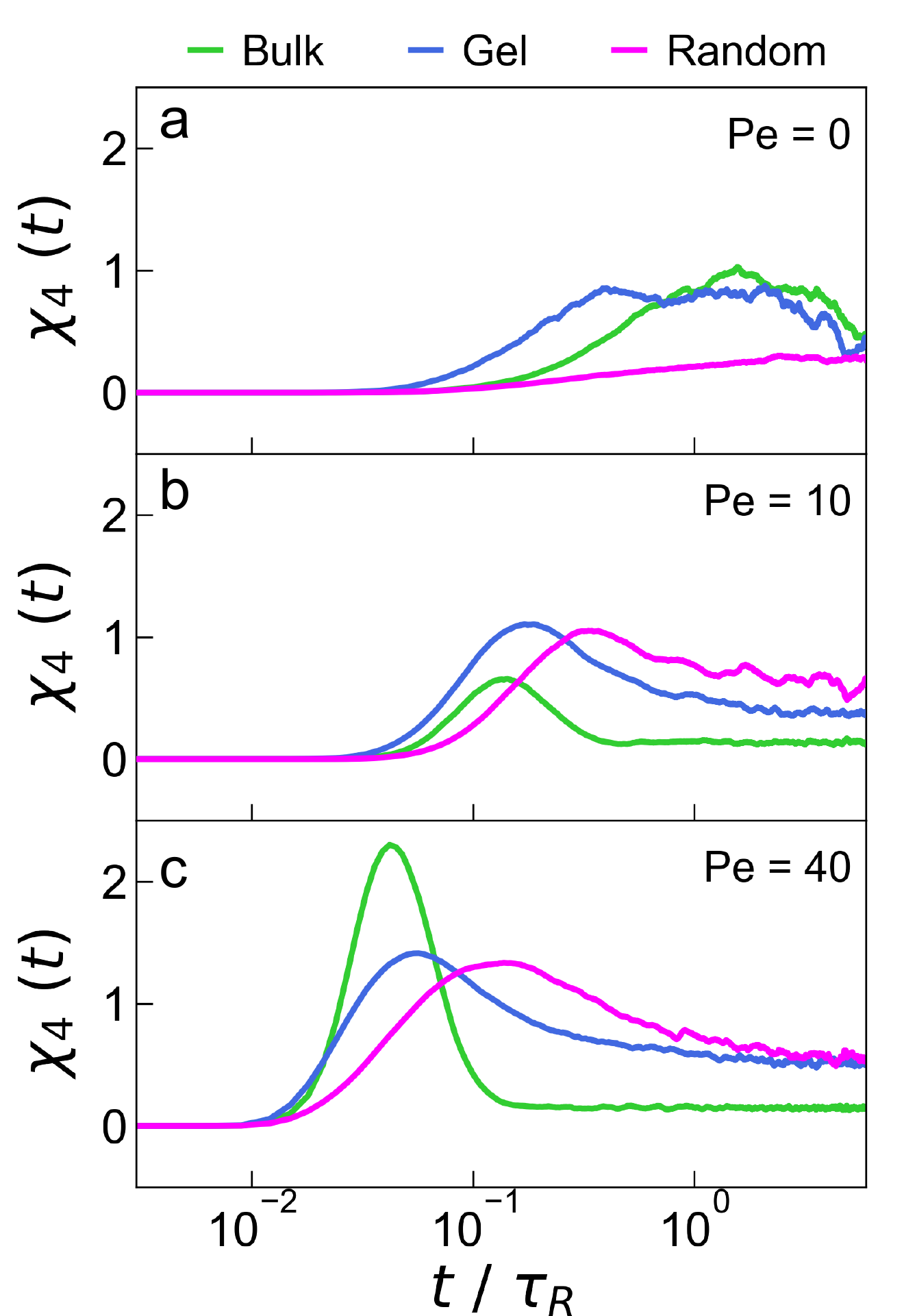}
\caption{The Dynamic susceptibility $\chi_4$ is measured in the three systems at $\rho = 0.87$ and at $\pecl = 0, 10, 40$ in panels (a), (b) and (c) respectively. The time corresponding to the peak of $\chi_4$ relates to the timescale of maximal dynamical correlation and the height corresponds to the number of particles involved in dynamic heterogenieties. Due to the large amount of sampling required, this data was measured using the smaller simulation box size.}
\label{figChi4}
\end{figure}

\textit{Random pinning ---}
For the random pinning case, an arbitrary configuration of particles is generated in the simulation box at the desired density. These particles then follow Brownian dynamics with the soft potential Eq. (\ref{eqsoft})  to eliminate any significant overlaps before the system is equilibrated with the WCA potential. A fraction of the particles are then chosen at random and frozen. This creates the random obstacles. This fraction is chosen such that the volume accessible to the free particles is the same in both the gel network and the random pinning systems (Fig. \ref{figObstacles}(c-d)).

The fixing of the accessible free volume of the mobile particles enables comparison of observables in both environments. Fixing this is a necessary step as the difference in structure between the gel network and the random pins could result in the free particles operating at two different effective densities in the case of the same number of frozen particles.

For this, the density of the gel network is held constant and the number of pinned particles is varied as a function of the total density $\rho$. The total free volume available to the mobile particles is determined by taking the Voronoi tessellation of the instantaneous configuration and associating with each particle the volume of its Voronoi cell $V^i_{\textrm{voro}}$. The sum of these volumes provides the total volume accessible to the free particles $\sum_{i}^N V^i_{\textrm{voro}}$, which are then averaged over ten independent simulation runs. This information is used to determine the fraction of particles to be pinned, such that $\sum_{i}^N V^i_{\textrm{\textrm{voro}}}$ for the free particles in the pinned system matches that of the porous gel network at the same density.

\textit{Lengthscales in the complex environments ---} The interplay between the lengthscale of the two environments and the persistent motion of the active particles will have a large impact on the dynamics. Therefore, to characterize the lengthscale the obstacles impose on the active particles we will use the pore \textit{chord length}. The chord length is a measure of the distance between two interfaces in a homogeneous phase of a heterogeneous system. A chord is defined as the distance between two interfaces in a heterogeneous system, where the chord lies wholly within one phase. The chord length distribution $p(\ell)$, defines the probability of finding a chord of length between $\ell$ and $\ell + d\ell$ within one phase. We characterize the environments in this work by the mean pore chord length $\langle L_c \rangle$. In practice, this is determined by measuring chords of varying lengths along each axis of the three dimensional sample that lie wholly within the pore phase \cite{whittle1999,testard2011}. The mean chord length was measured and averaged for six independent configurations for each environment species. The gel networks have a mean pore chord length $\langle  L_c \rangle  = 6.62\sigma$, whereas for the random pinning system $\langle  L_c \rangle = 3.24\sigma$, almost half that of the gel system. The difference in this measurement derives from arrangements of the particles comprising these structures: in the case of the gel, particles are arranged locally into dense branches and therefore the branches provide the relevant lengthscale in this system. Conversely, in the random system the particles are arranged in a non-overlapping random configuration, and the surrounding free space is then dependent on the shorter lengthscale of the average particle separation.

\subsection{Dynamical analysis}
The addition of obstacles into dense fluids greatly influences the dynamics, and in some cases a system may become arrested. The structural relaxation time $\tau_{\alpha}$ provides a useful metric which one can use to understand the variation of timescales across different state points and environments.

The relaxation time $\tau_{\alpha}$ is determined via the self part of intermediate scattering function:
\begin{equation}
F_{\mathrm{s}}(k, t)=\frac{1}{N}\left\langle\sum_{j=1}^{N} \exp \left[i \vec{k} \cdot\left(\vec{r}_{j}(t)-\vec{r}_{j}(0)\right)\right]\right\rangle
\end{equation}
\noindent
where $\vec{k}$ is the wavevector $k=|\vec{k}|$, taken as $2\pi$. We define $\tau_{\alpha}$ as $F_s(k = 2\pi, \tau_{\alpha}) = e^{-1}$. Here the index $j$ runs over all the particles.

The persistent motion of active particles induces clustering and aggregation at boundaries. This will likely cause density fluctuations where some fraction of particles are in dense and crowded regions while others are in locally dilute regions. To quantify the degree to which these variations are taking place within different environments, we use the four-point dynamic susceptibility $\chi_4$. To calculate $\chi_4$, we follow the methodology of La{\v c}evi{\'c} \textit{et al.} \cite{lacevic2003}. For this, one must first define an overlap function $w\left(\left|\mathbf{r}_{j}(0)-\mathbf{r}_{i}(t)\right|\right)$, where $i$, and $j$ are particle indices. This measures the degree of spatial similarity between configurations of a system as a function of time. The overlap is unity inside a region  $\left|\mathbf{r}_{j}(0)-\mathbf{r}_{i}(t)\right| \leq a$ and $0$ otherwise, where $a = 0.3 \sigma$. The fraction of overlapping regions in a system of particles at times $0$ and $t$ is given by:

\begin{equation}	
Q(t)=\frac{1}{N} \sum_{j=1}^{N} \sum_{i=1}^{N} w\left(\left|\mathbf{r}_{j}(0)-\mathbf{r}_{i}(t)\right|\right).
\label{eqOverlap}
\end{equation}

The fluctuation of $Q(t)$ then defines $\chi_4$. This quantity is a susceptibility and measures dynamic heterogeneity \cite{berthier}.

\begin{equation}
\chi_{4}(t)=\frac{V}{N^{2} k_{B} T}\left[\left\langle Q^{2}(t)\right\rangle-\langle Q(t)\rangle^{2}\right].
\label{eqChi4}
\end{equation}

\begin{figure*}
\centering
\includegraphics[width=0.68\linewidth]{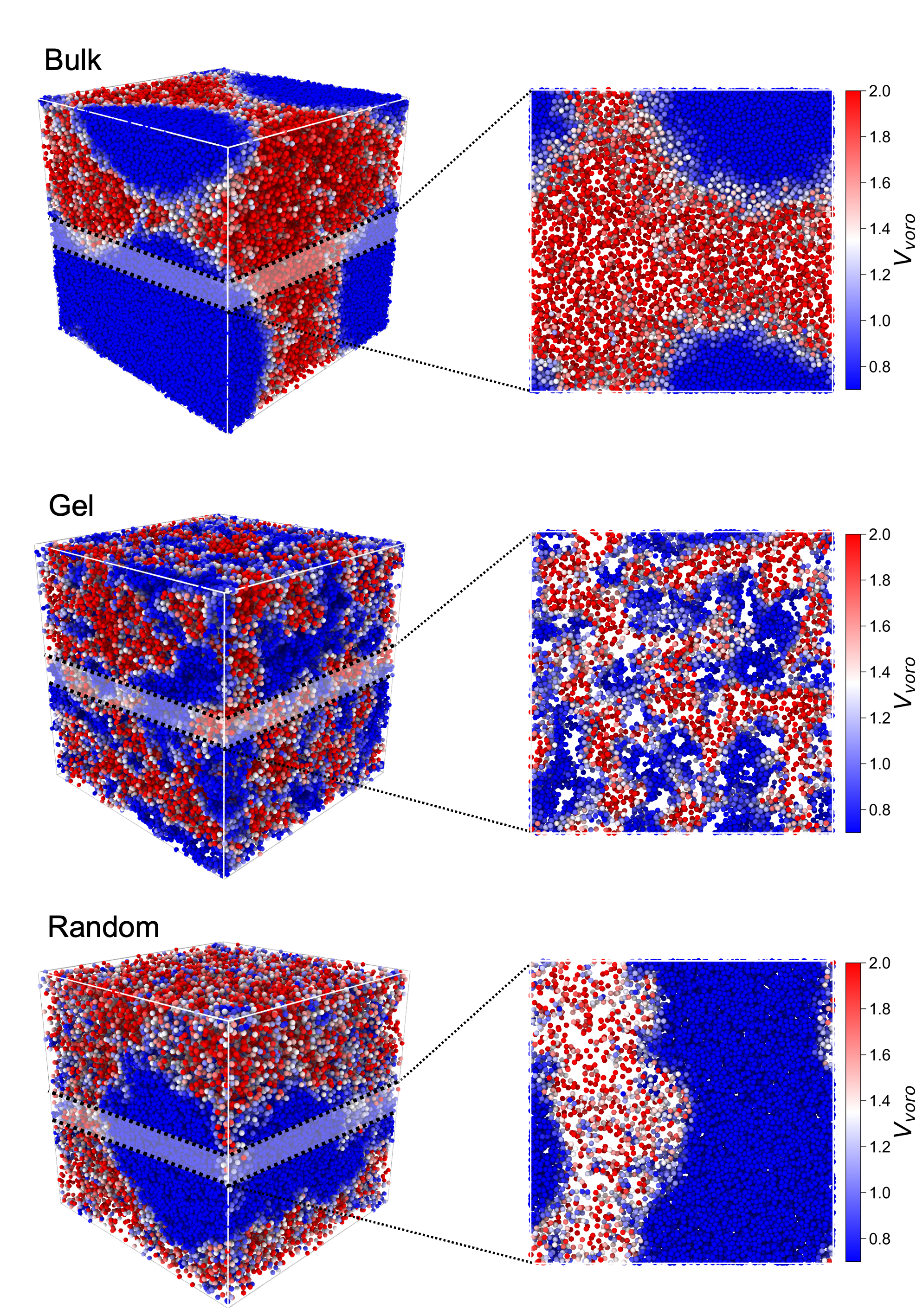}
\caption{(\textit{Left}) 3d Snapshots of the three systems at $\pecl = 100$, particles are coloured by their Voronoi volumes ($V_{\textrm{voro}}$), obstacles are not rendered. (\textit{Right}) A slice through each system (depth $4\sigma$).}
\label{fig3DSnapshots}
\end{figure*}

\section{Results}
\label{sectionResults}

Our results section is organized as follows. We first characterize the dynamical behavior at low activity characterized by the P\'{e}clet number. We then move on to moderate activity, where significant structural changes are observed, related to motility induced phase separation. These we characterize with a number of measures, one--body measures such as the Voronoi volume associated with each particles, and many--body properties, accessed with the topological cluster classification \cite{malins2013tcc}. Finally, we probe the case of very high activity, where we observe a re--entrant mixing.

\subsection{Low $\pecl$: crystal suppression and heterogeneous dynamics}

\textit{Structural relaxation ---} As mentioned in the introduction, active systems can undergo a localization transition in the presence of quenched disorder. This phenomenon is often only present in the homogeneous phase i.e. for activity below that associated with the boundary of motility induced phase separation. To assess the extent to which complex environments impose localization on active Brownian particles, we measure the structural relaxation time ($\tau_{\alpha}$). This enables us to to determine the regimes in which these systems become arrested. Figure \ref{figTauAlpha} shows the variation of the structural-relaxation time in the bulk, gel network and random pinning systems as a function of $\pecl$ and total number density $\rho$.

In the bulk system (Fig. \ref{figTauAlpha}-bulk), the particles will crystallize for state points that fall below the freezing line. This crystal regime was studied previously from which the data for this phase boundary originates \cite{moore2021}. Outside of the crystalline regime, the active particles exhibit a monotonic decrease in $\tau_{\alpha}$ as $\pecl$ rises for a given density, which is qualitatively similar to a temperature increase in passive systems \cite{malins2013fara}.

With the addition of a complex environment, there is an emergence of slow dynamics distinct to that of the crystal in the bulk system. Figure \ref{figTauAlpha}-gel and Fig. \ref{figTauAlpha}-random show the variation in $\tau_{\alpha}$ for the gel network and the random pins respectively.  Comparing the two panels, we can see that the two systems operate on significantly different timescales, with the system with random obstacles exhibiting slower dynamics than the gel network for the majority of state points considered. Both the gel network and the random pinning systems exhibit structural relaxation times in excess of $10^{5}\tau_R$ at high densities, plotted as the darkest blue contour. These are state points which will not relax within the maximum run time of the simulations and are thus for the purposes of this work, we define these as arrested. In this arrested regime, active particles are localized due to the constraints imposed by the gel or the pins, even at P\'{e}clet numbers of the order of $10$ for dense ensembles in the random pins.

\textit{Dynamic Susceptibility ---} The interplay of the complex environment and the self propulsion will cause a degree of dynamical correlation in these systems, as clusters of active particles become aligned or absorbed at the boundaries. The dynamic susceptibility $\chi_4$ provides a measure of this, it is plotted for the three systems in Fig. \ref{figChi4}. The dynamic susceptibility manifests a peak at the timescale for which the particle dynamics are maximally correlated on the chosen lengthscale $a$ (Eq. \ref{eqChi4}). This measure shares some similarities with the structural relaxation time. For example, in the passive systems ($\pecl=0$), the positions of the peaks indicate the same relationship as measured in $\tau_\alpha$ in Fig. \ref{figTauAlpha}, with the gel system being the fastest, followed by the bulk and then the random system coming significantly later. At low activity ($\pecl=10$), these peaks become more clearly defined and move to shorter times. Furthermore the relative positions of these peaks switch, with the dynamic correlations in the bulk system  happening on shorter timescale than the complex environment systems. Finally, at higher $\pecl$ (the regime of motility-induced phase separation), the dynamic correlations are appreciably more prominent in the bulk system. Conversely, it is clear that for the gel and random system, $\chi_4$ does not fully relax on the timescales we probe, as the environment enforces some degree of dynamical correlation over long timescales.

\subsection{Intermediate $\pecl$: MIPS in random environments}

\begin{figure}
\centering
\includegraphics[width=\linewidth]{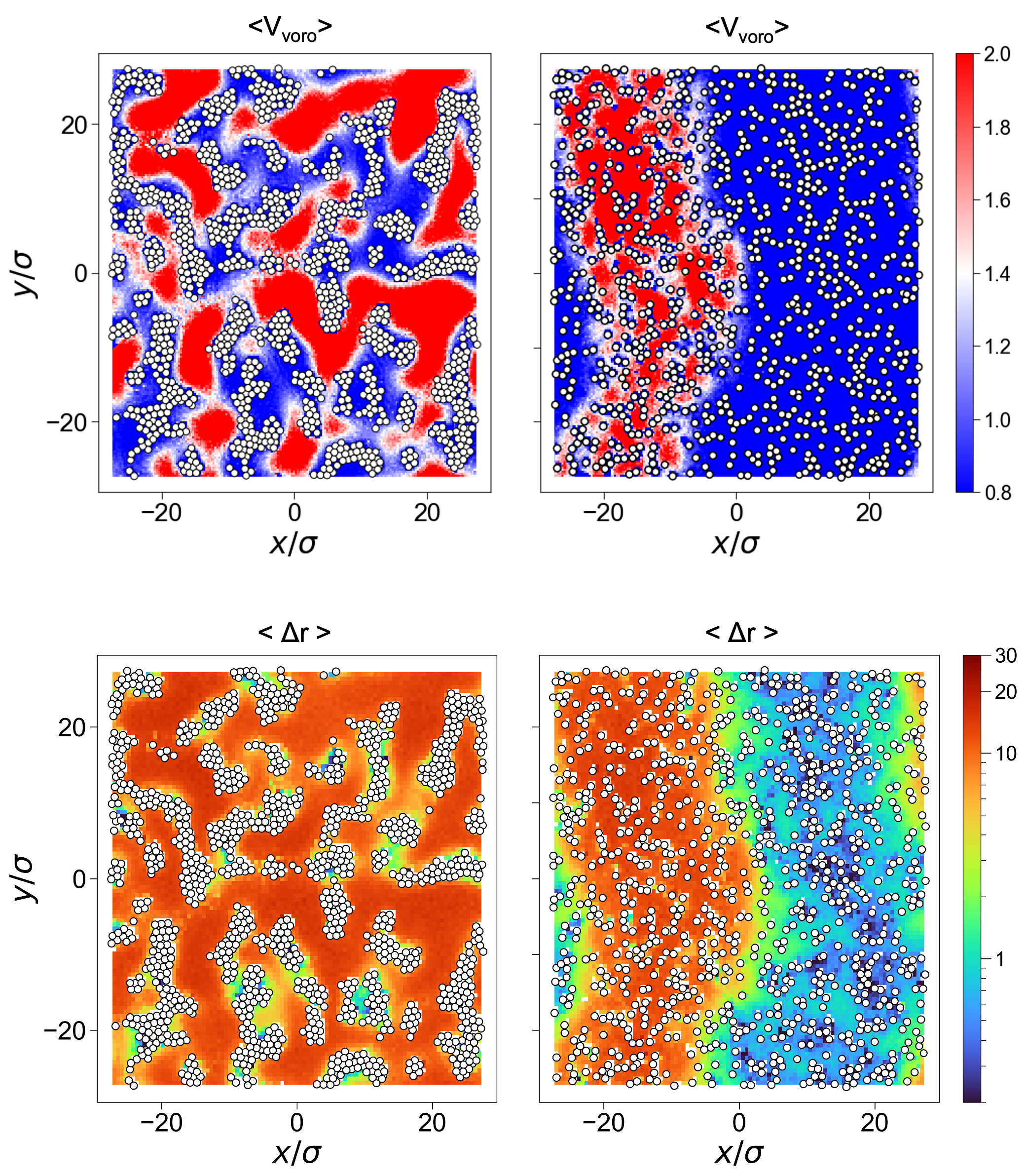}
\caption{Heat map of Voronoi volumes and single particle displacements in the gel network (left), and random pinning (right) systems respectively. Note that displacements are measured over $\Delta t = 6 \tau_R$. All systems are at $\rho=0.87$, and  $\pecl=100$. %various $\pecl$ (see figure legend).
}
\label{figheatmaps}
\end{figure}

\begin{figure*}
\centering
\includegraphics[width=\linewidth]{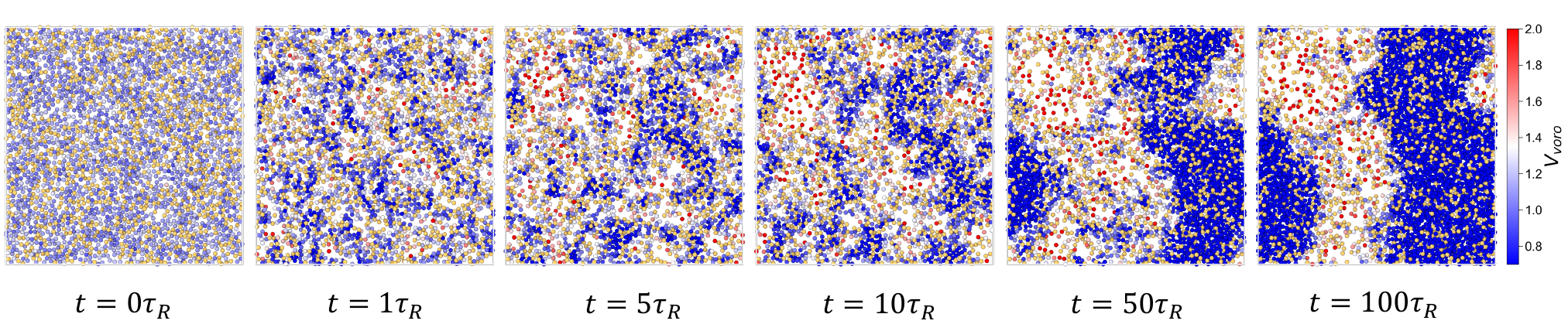}
\caption{Time--evolution of the pinned system undergoing MIPS. The state point is that of Figs. \ref{figheatmaps} (right), $\rho=0.87$, and  $\pecl=100$. The color map represents the Voronoi volumes of the active particles and the pale yellow particles are the pinned particles. Each snapshot is taken at the time noted underneath from the start of active motion. Data are taken from a slice of depth $\sigma$.}
\label{figNucleation}
\end{figure*}

\begin{figure*}
\centering
\includegraphics[width=0.8\linewidth]{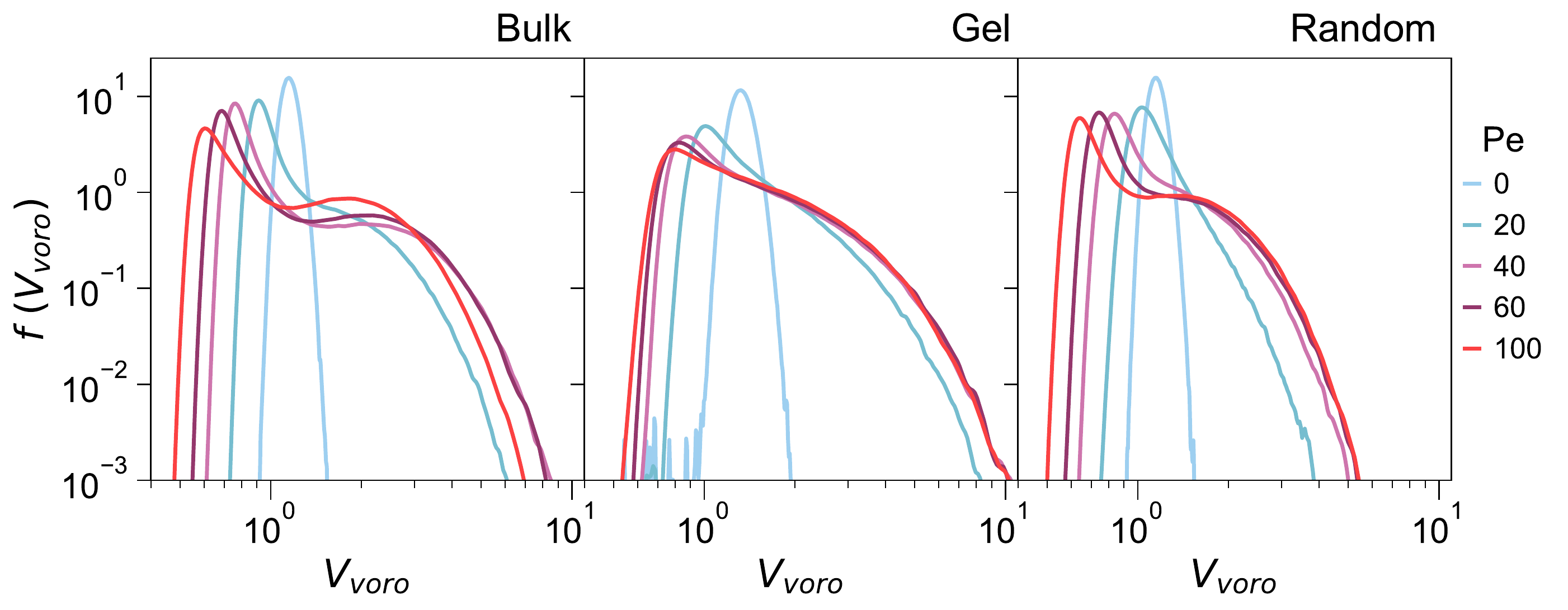}
\caption{Probability density of the Voronoi volumes ($V_{\textrm{voro}}$), for the bulk, gel, %porous network, 
and random pinning systems respectively. All systems are at $\rho=0.87$, and various $\pecl$ (see figure legend).}
\label{figVoroPDF}
\end{figure*}

\begin{figure}
\centering
\includegraphics[width=\linewidth]{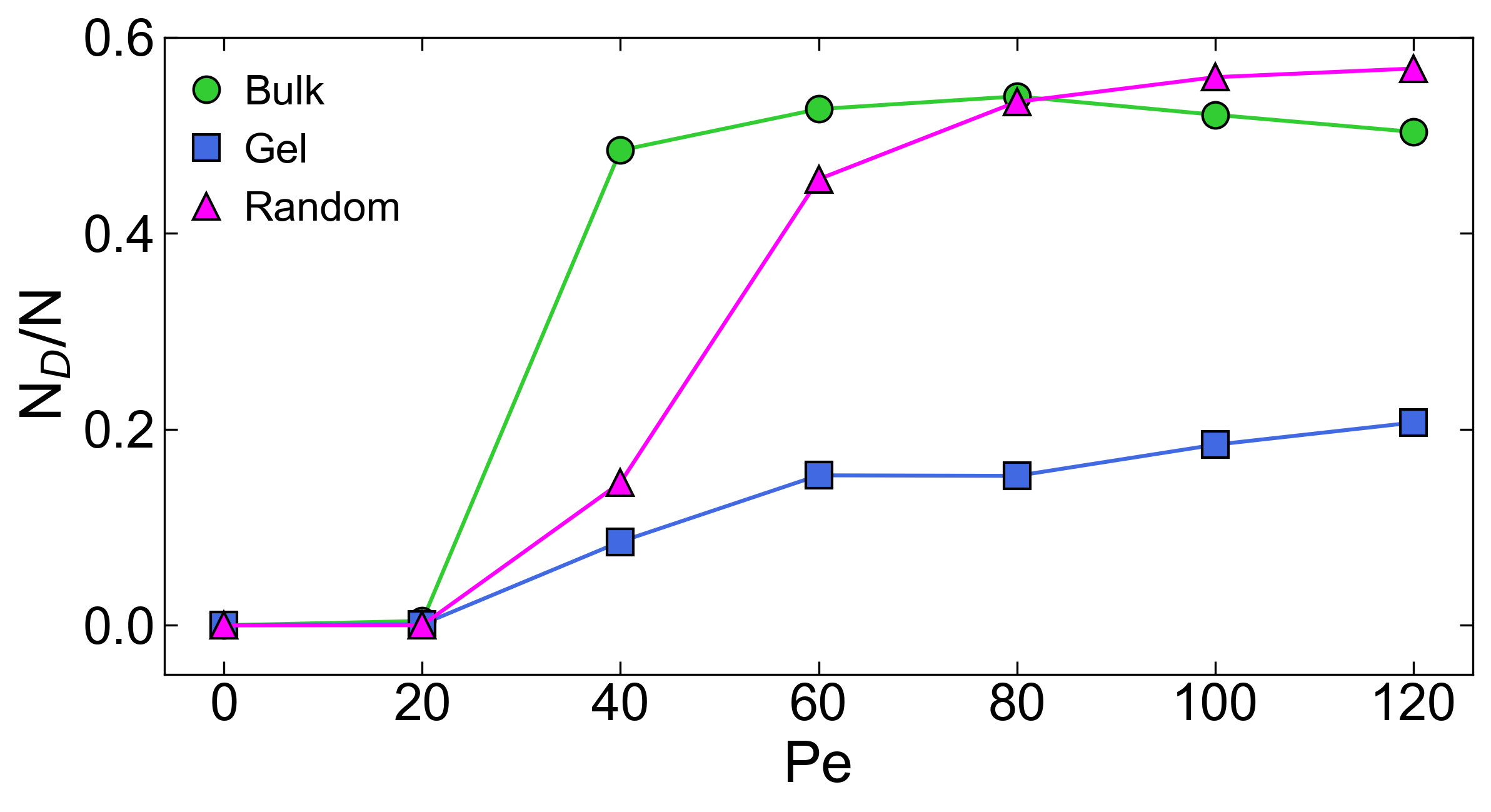}
\caption{The fraction of locally dense particles $N_D / N$.  Particles are considered locally dense if $V_{\textrm{voro}} < 0.8$.}
\label{figLocalised}
\end{figure}

From the results presented in the previous section it is clear that the different confinement types cause significant and varying perturbations to the dynamics of active systems at low $\pecl$. What remains unclear, is the mechanism by which active particles overcome localization due to the environment, and what behavior these particles exhibit when driven towards higher P\'{e}clet numbers, particularly in the regime of motility induced phase separation (MIPS). Compared to the lengthscale of MIPS, we know that the complex environment restricts the active particles to motion over shorter lengthscales; from the chord length measurements, the confining lengthscale is $6.62\sigma$ in the gel and $3.24\sigma$ in the case of random pinning.

We use the individual particle Voronoi volumes ($V_{\textrm{voro}}$) as a measure of the local density to provide insight into the relationship between self-propulsion and complex environments. Figure \ref{fig3DSnapshots} displays representative snapshots of the bulk, gel network and random pinning systems in the steady-state at an %high 
activity of ($\pecl = 100$), alongside a slice through each system. In the bulk system (top panel) the particles have undergone motility-induced phase separation. Within the large dense region are particles (colored dark blue, which have $V_{\textrm{voro}} \leq 0.8$ in Fig. \ref{fig3DSnapshots}), clustering due to their persistent motion, which are surrounded by a gas of active particles at a lower density.

Figure \ref{fig3DSnapshots} (middle panel) displays the behavior of active particles in the gel at high $\pecl$. The structure of the gel is non-trivial featuring extreme variations in surface curvature and channel width. It is clear that this has a strong impact on the dynamics, with a distinct pattern emerging which is clearly dependent on the gel structure. In particular MIPS leads to a highly complex structure, with some pores being filled by the slow/dense phase and others by the fast/less dense phase. The location of the regions which acquire a low or high density under MIPS is fixed by the initial configuration, with independent runs always exhibiting the same demixing pattern.

The influence of the random pinning (bottom panel) on the active particle dynamics at high $\pecl$ is distinct from that of the bulk and gel systems. Here, the particles phase separate into a large cylindrical droplet somewhat akin to the bulk system. However, this droplet has random pins throughout its structure holding the dense phase in space. Furthermore, the presence of the random pins brings about an unexpected result: the arrangement of the random pins pre-defines the location in which the dense MIPS phase will form. For the same configuration of pinned particles, but distinct alternate initial positions and velocities; MIPS occurs in roughly the same region of the system, with only small fluctuations from run to run. Therefore, we believe that the location of the MIPS is somehow encoded in the initial positions of the randomly pinned particles.

The three-dimensional snapshots in Fig. \ref{fig3DSnapshots} provide a good indication of the phase behavior of these systems. However, they do not contain any information on the stability of the dense phase. Therefore, in Fig. \ref{figheatmaps} we plot the average Voronoi volume $\langle V_{\textrm{voro}} \rangle $ as a function of space in both the gel network and random pinning systems. These plots give information regarding the way in which each system phase separates. For the gel, the inclusion of the network into the space does not allow for the formation of a single dense droplet as seen in bulk systems experiencing MIPS. Instead, the structure of the gel determines the locations in which the system will phase separate. Given the disordered nature of the gel, there are sections where the pores are more constricted, have tighter curvature or are less connected; these are the locations that will trap active particles. The spatial distribution of $\langle V_{\textrm{voro}} \rangle $ in the random pinning system tells an alternate story. Active particles in this system at sufficient $\pecl$ will form a large droplet around a subset of pinned particles. The random pins in this droplet keep it stable over long time periods in a steady state where particles are exchanged between the droplet and the surrounding active gas.

The Voronoi volumes give information into the phase separation of these systems but not into the transport dynamics of the individual particles in this environment. Some insight into this aspect can be gained by looking at the average single particle displacements $\langle \Delta r \rangle$ for the same system. The distributions of $\langle \Delta r \rangle$ are plotted in Fig. \ref{figheatmaps} for displacements over the time period $t=6\tau_R$, which is a time period of the order of $\tau_{\alpha}$ in the passive bulk system at this density. For the gel system, these displacements are largely uniform in the wider channels. Close to the surfaces of the gel particles the displacements are significantly less, indicating shorter movements along the surface. Furthermore, there are several locations where particles are localized with displacements less than the particle diameter. These are all located in regions of high surface curvature.

\begin{figure*}
\centering
\includegraphics[width=\linewidth]{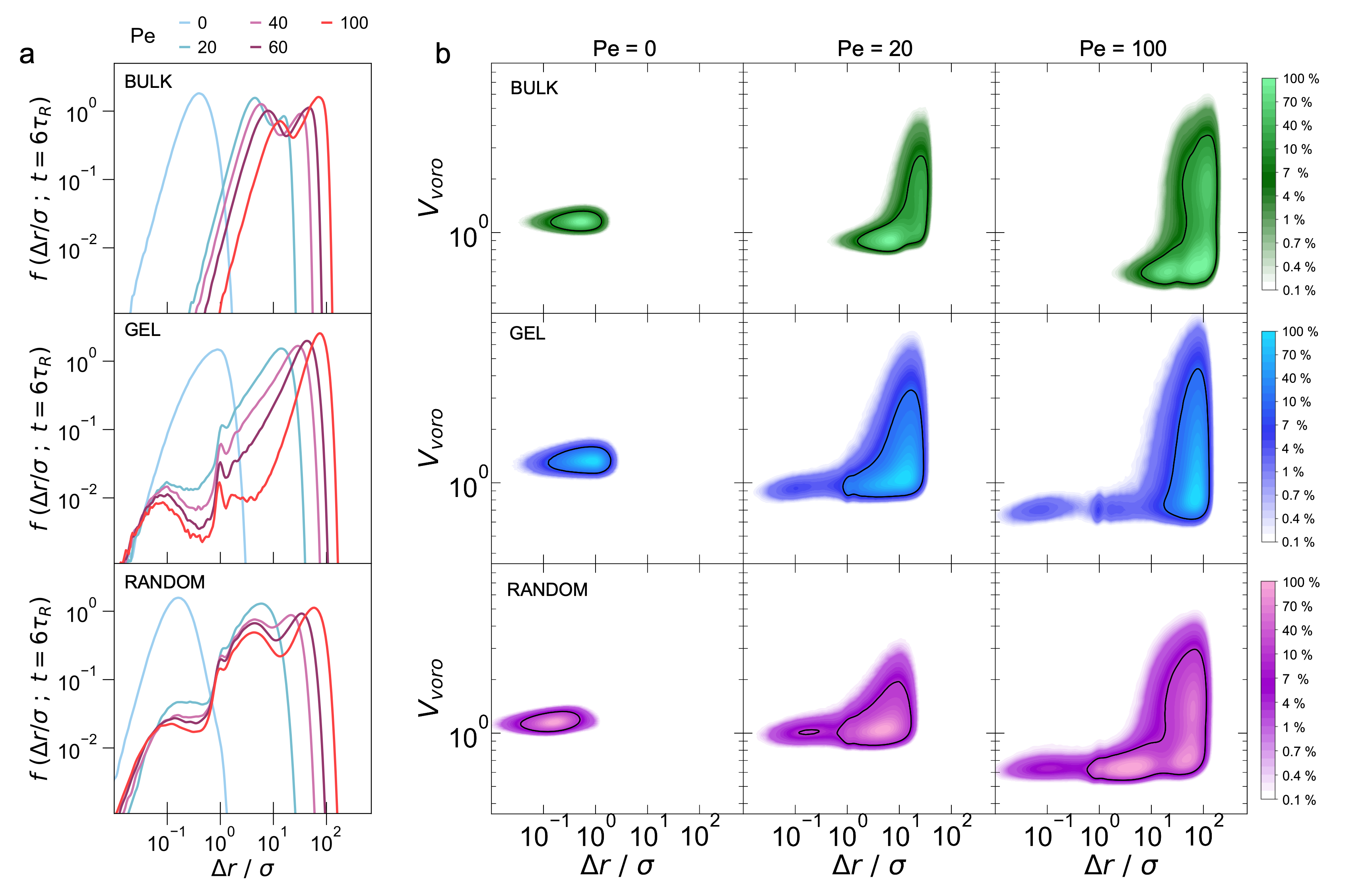}
\caption{(a) Probability density of the Voronoi volumes ($V_{\textrm{voro}}$), for the bulk, porous network, and random pinning systems respectively. All systems are at $\rho=0.87$, and various $\pecl$ (see figure legend). (b) Correlation of the Voronoi volumes and the single particle displacements of motile particles in the gel network, random pins and the bulk system. All systems are at a density of $\rho = 0.87$, and plotted for $\pecl = $ 0, 20, and 120. Black lines contain 90\% of data. Colours show contour levels below which the indicated percentage of the data will lie beneath.}
\label{figCorrelation}
\end{figure*}

For displacements in the random system, there are four identifiable populations of particles, each with its distinct environment. The first is the group of localized particles. These are particles that have become trapped between the random pins and other particles in the dense phase and are recognizable as the dark spots dispersed through the dense phase. These are surrounded by the second group of particles that are not localized but remain trapped within the dense phase and are moving very slowly ($\Delta r < \sigma$). Beyond this are particles in the interface, which undertake mid-range displacements. Finally, outside the dense phase is the active gas where particles are completing relatively large and uniform displacements.

\vspace{10pt}
\textit{Time-evolution of MIPS in random pinning --- }
Our protocol enables us to investigate the process by which the system undergoes MIPS. To this end, we show a time--sequence of snapshots of a pinned system undergoing MIPS. In Fig. \ref{figNucleation} we show the formation of MIPS, starting from the passive WCA system as described in section \ref{sectionModel}. Here we consider the same state point as in the previous figures  ($\rho=0.87, \pecl=100$) and plot the Voronoi volumes as in Fig. \ref{figheatmaps} (top row). The time--evolution of this data is also shown in the supplementary movie.

At time $t=0$, we see that the system is largely uniform in density. Visual inspection indicates the following. Even at quite small times ($t\lesssim \tau_R$), there are fluctuations in density which are small, both in the change of local volume per particle and also in their spatial extent. Over time, the larger fluctuations (in terms of their spatial extent) appear to grow at the expense of smaller fluctuations and even at the quite short time of $10 \tau_R$, the pattern seems to be largely fixed. Longer times correspond to an increase in density difference between the particle--rich regions (blue) and the more dilute regions. During this time regime, the interface between these regions becomes better defined. While we have been able to observe the time-evolution of MIPS under random pinning, exactly what causes the spatial distribution of the MIPS phases and how this is encoded in the random pins seems to be a challenging problem for the future.

\vspace{10pt}
\textit{Density fluctuations as a function of $\pecl$ ---}
So far we have seen examples of how these systems phase separate at $\pecl\approx100$, now we will look at how the local density fluctuates in these systems for different $\pecl$. The probability density of the Voronoi volumes for each system % determined via kernel density estimation and 
are shown in Fig. \ref{figVoroPDF} for various $\pecl$. In the absence of activity ie. $\pecl=0$, all systems feature an approximate Gaussian distribution. Common to all the environments there is the fact that the addition of activity produces a shift in the peak towards smaller volumes and a broadening of the tail of the  distribution towards larger volumes. In the bulk system (Fig. \ref{figVoroPDF}-bulk) the distributions feature a non-monotonic trend in the spread. First increasing as a function of $\pecl$ up to a maximum at $\pecl = 60$, before decreasing. This is the first sign of re-entrant MIPS mixing, %behavior, 
which will be the focus of Section~\ref{sectionMotility}.

For the gel network (Fig. \ref{figVoroPDF}-gel) and the random pins (Fig. \ref{figVoroPDF}-random), this broadening is monotonic, with the gel network covering a wider range of volumes. However, for the random pins and the bulk systems we observe the emergence of twin-peaked distributions at higher activity as a result of phase separation. For both the gel and the random pins, the influence of the complex environment leads to a splitting of the active particles into more than one population, with a proportion of active particles aggregating or becoming localized because of interactions with the environment.

The persistent motion of active particles induces symmetry breaking that causes them to aggregate at surfaces and walls. In these systems the surfaces could be the edge of a MIPS dense phase, the surface of the gel network, or a dense cluster of pins. These surfaces collect active particles. To determine how the environment structure affects the collection of particles at surfaces, we count the number of particles located in locally dense regions $N_D$, defined for particles where $V_{\textrm{voro}} < 0.8$. In Fig. \ref{figLocalised} we plot the fraction of localised particles $N_{D}/N$ as a function of $\pecl$.

\vspace{10pt}
\textit{Active transport in complex environments ---}
We have seen so far that with a progressive increase in $\pecl$, all three systems undergo dramatic changes in terms of local density. We have also seen that at high $\pecl$ the average single particle displacements reveal information concerning %in to 
the interaction the active particles have with their environments. Like with the Voronoi volumes, we plot the probability densities of active particles in the three systems for various $\pecl$ [Fig \ref{figCorrelation}(a)]. At first glance, it is clear that for all systems the particles complete larger displacements as $\pecl$ is increased. Moreover, these distributions are highly featured and reveal a great deal of information to the behaviour and interactions of the active particles.

In the bulk system [Fig. \ref{figCorrelation}(a-bulk)], the $\Delta r$ distribution is a single peaked at $\pecl$=0. As $\pecl$ increases, this distribution shifts to higher displacements and we observe the growth of a second peak; indicating the presence of MIPS in the system. With a further increase of $\pecl$ there is a continuous transition between the relative heights of the peaks as the fraction of particles in the dense phase grows. This is corroborated by the growth of regions of locally dense particles over this range of $\pecl$ observed previously in Fig. \ref{figLocalised}.

\begin{figure*}
\centering
\includegraphics[width=0.8\linewidth]{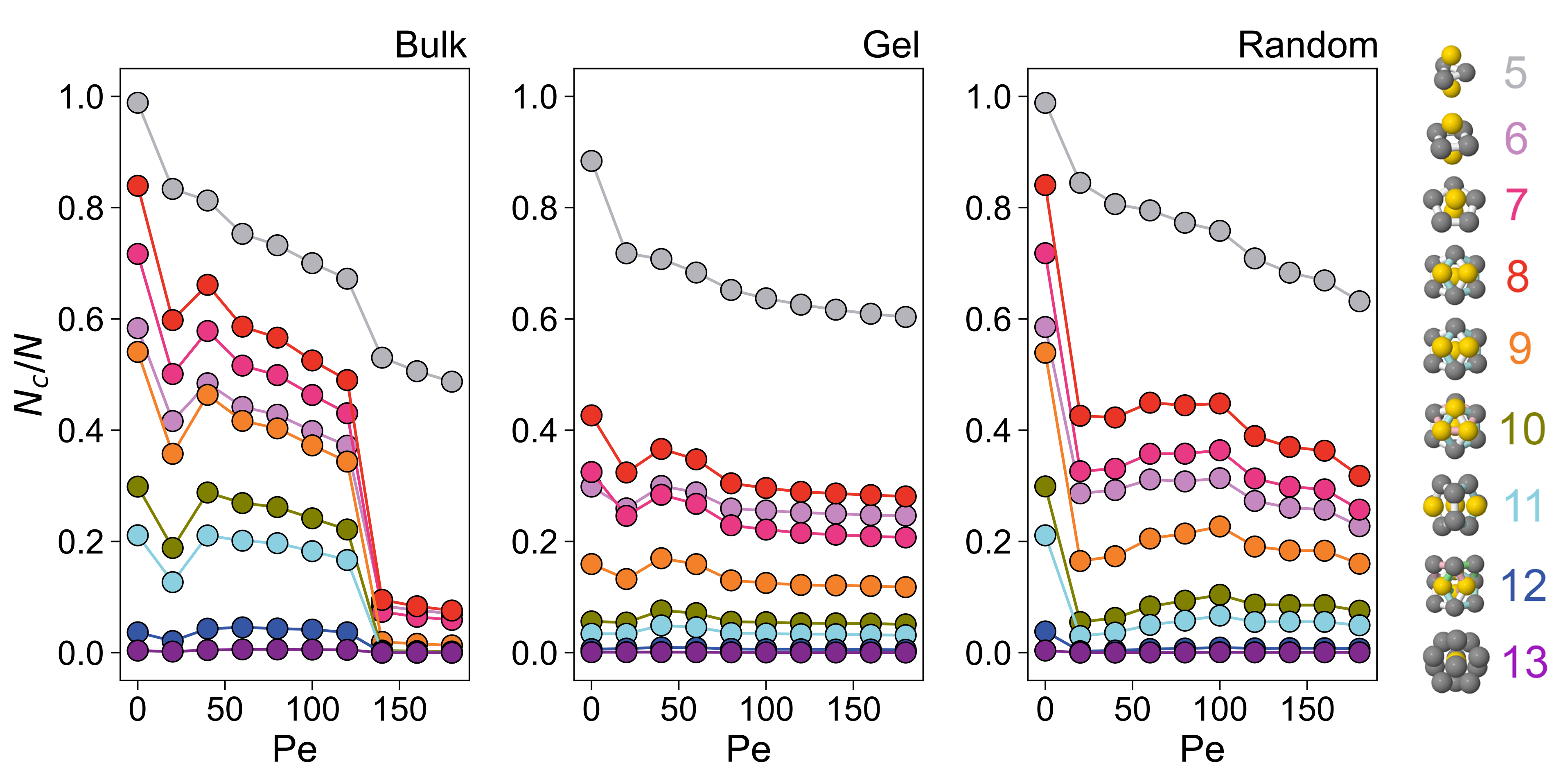}
\caption{Higher--order structural analysis using the topological cluster classification. Here particles in specific local environments corresponding to clusters of 5-13 particles, and the hexagonal close-packed and face centered cubic crystals are considered. The number of particles in each environment $N_c$ is then plotted as a function of $\pecl$ for the three geometries under consideration, bulk, gel and random systems at $\rho=0.87$. Colors of the lines and data points correspond to the clusters depicted in the legend. The colors of the particles in the renderings in the key correspond to geometric properties of the clusters. In particular, the grey particles are in 3,4 or 5-membered rings and the yellow particles correspond to so-called ``spindles'' \cite{malins2013tcc}.}
\label{figTCClabelled}
\end{figure*}

\begin{figure*}
\centering
\includegraphics[width=\linewidth]{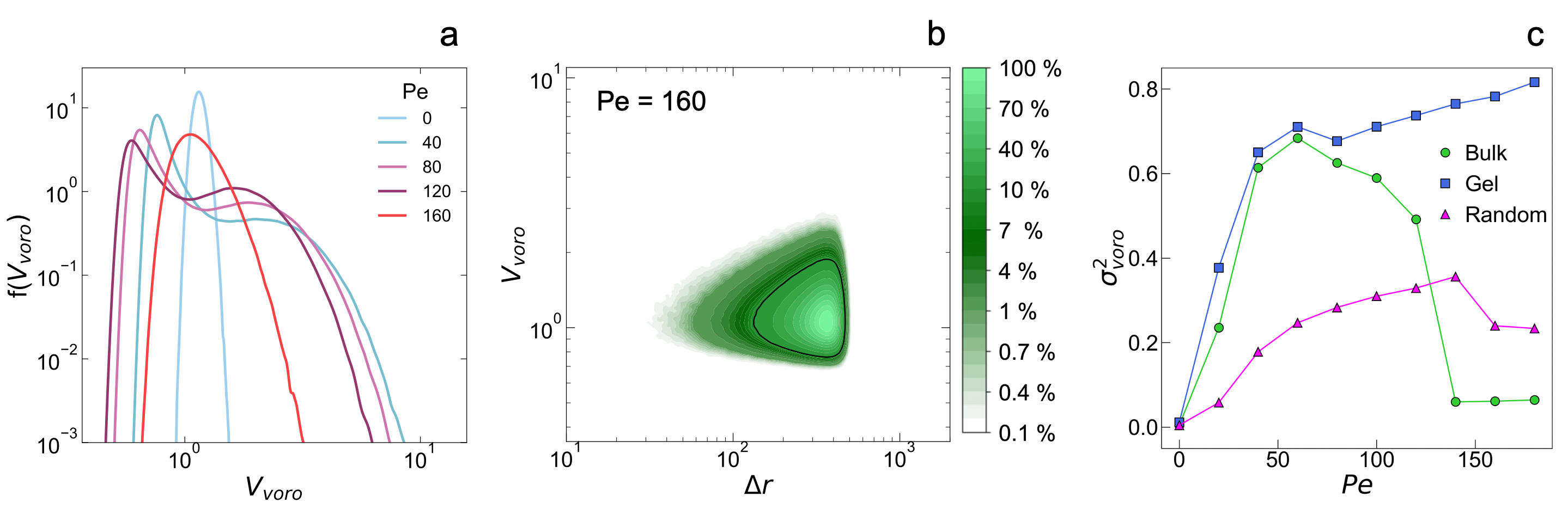}
\caption{(a) Probability density of the Voronoi volumes in the bulk system at various $\pecl$. As $\pecl$ increases the distributions show first one, then two populations, and then a re-entrant single phase at high $\pecl$. (b) Correlation of $V_{\textrm{voro}}$ and $\Delta r$ in the bulk system at $\pecl=160$ shows a single population. (c) Variance of the distribution of Voronoi volumes in the porous network ($\sigma^{2}_{voro}$), random pinning and the bulk systems as a function of $\pecl$, all at density $\rho = 0.87$.}
\label{figReentrant}
\end{figure*}

The displacements for the gel network and the random pinned system are plotted in Fig. \ref{figCorrelation}(a-gel) and Fig. \ref{figCorrelation}(a-random) respectively The distributions of both of these systems show the splitting of a single population into two or more populations with the progressive increase of $\pecl$. The first of these is the emergent peak at very small displacements $\Delta r / \sigma \sim 10^{-1}$. These particles %are 
move only a small fraction of their diameter and have become localized as a result of the interplay between their activity and the environment.

Interestingly, both systems in complex environments [Fig. \ref{figCorrelation}(a-gel/random)] feature a strong peak at $\Delta r / \sigma = 1$, with some smaller features at subsequent integer displacements. Examination of the average displacements in Fig. \ref{figheatmaps}, shows that they %these displacements 
are located along the surfaces of the gel network and in small pockets of lower density in the dense phase of the pinning system. The location of these displacements makes it clear that they are a result of particle re-arrangements at the obstacle interface, and in dense particle clusters. For the gel network [Fig. \ref{figCorrelation}(a-gel)] at $Pe > 0$, the remaining particles are in a single large population, moving comparable distances to particles in the bulk system. However, %Whereas, 
in the case of the random pinned system [Fig. \ref{figCorrelation}(a-random)], at $\pecl \geq 40$ there is a splitting of larger displacements across two length scales, one at the interface and the second in the active gas.

Thus far we have primarily considered two observables $V_{\textrm{voro}}$ and $\Delta r$, both of which tell part of the story. These two observables can be correlated to complete this picture, the result of this is plotted in Fig. \ref{figCorrelation}b. The result is a series of contours stacked logarithmically, each layer denotes the level at which a percentage of the data lies below. These plots provide some insight into the population splitting phenomena we have seen so far. We see that the combination of confinement and activity create a subpopulation of particles that are arrested and have very little free space,  a feature not found in the bulk system. Interestingly, this arrested group is relatively larger in the random pinning system. For the systems at high $\pecl$ in Fig. \ref{figCorrelation}(b), the particles that have $V_{\textrm{voro}} < 0.8$ and in some form belong to dense clusters, these particles still cover a wide range of displacements. This is likely a combination of two phenomena, first is that of the particles that have been localized over a longer time frame. These are particles that are not moving and have very little free space. The other case, is that of particles which have been mobile but have very little space. These will be particles that have moved from a position at a previous time and are now located in a dense cluster or at the object interface.

\subsection{Local structure}
\label{sectionLocalStructure}

So far, we have focussed mainly on one--body structural properties via the Voronoi volumes. It is instructive when studying amorphous systems to consider higher--order structural correlations, as a precise means to probe smaller changes in the local structure. To this end, we use the topological cluster classification (TCC) \cite{malins2013tcc}. The TCC identifies local environments whose bond topology is identical to that of small minimum energy clusters of a suitable reference system, here the Lennard--Jones model. Some of us have shown that this is appropriate for passive WCA particles \cite{taffs2010} and also have investigated the effect of activity on a similar (bulk) system \cite{dougan2016,moore2021}. Here, we consider clusters of differing sizes (5-13 particles) as indicated in Fig. \ref{figTCClabelled}. The number of particles in each of these clusters $N_c$, scaled by $N$ is then plotted as a function of $\pecl$ for $\rho=0.87$. We further consider the influence of the local environment where $N_c$ is the number of particles participating in a particular cluster. To identify the cluster population we need the bond network, which we identify with a modified Voronoi decomposition (specifically we set the Voronoi parameter of Ref.~\cite{malins2013tcc} to $f_c=0.82$). Now we are interested in the active particles, but in order to identify the clusters, we include the immoblized particles in the bond network. We then run the TCC, but when analyzing data, we only consider active particles.

Figure \ref{figTCClabelled} shows a common trend between the bulk, gel and random environments: the equilibrium local structure population is disrupted by increasing $\pecl$. Both of the confined environments show cluster populations that are lower compared to the bulk case. The random environment, where the position of the obstacles is taken from equilibrium configurations, at $\pecl=0$ has the same population numbers as the bulk case, but as soon as activity is switched on all cluster populations fall more abruptly compared to the bulk case. The first rapid decrease in local structure at $\pecl\simeq 30$ coincides with the formation of MIPS. Unlike the gel and random pinning systems the bulk case experiences another large drop in local structure population for $\pecl\simeq 140$. As we will see in the next section, this drop corresponds to a re-entrant mixing. Local structures are present in lower population in active systems than in similar equilibrium systems as has been seen previously \cite{dougan2016}. However, it is possible that phase separation or the influence of the environment could influence this.

\subsection{Motility-Induced Mixing}
\label{sectionMotility}

In addition to the phase behavior discussed thus far, for our purposes, there is one more regime to be considered. At very high $\pecl$, ABPs will transition from a demixed state due to motlity-induced phase separation to a mixed state, ie a homogeneous active fluid. A similar behavior has been observed previously in Ref.~\cite{bialke2013,stenhammar2014} in bulk systems.  We confirm these observations in the bulk case. The presence of the transition is apparent in the probability density of $V_{\textrm{voro}}$ in the bulk system [Fig. \ref{figReentrant}(a)]. These distributions show two populations at  intermediate $\pecl$, but a single population at $\pecl=0$ and at $\pecl > 120$. Looking at the correlation of $V_{\textrm{voro}}$ with $\Delta r$ confirms the single re-entrant phase at high $\pecl$ [Fig. \ref{figReentrant}(b)].

To study the re-entrant MIPS behaviour in systems with quenched disorder in Fig. \ref{figReentrant}c we plot the variance of the probability density of $V_{\textrm{voro}}$ as a function of $\pecl$: non monotonic behaviour in this quantity signals a re-entrant phase. Looking at this data we can see that for the bulk and random systems, as $\pecl$ increases, the variance also increases up to a maximum; beyond which it decays to an intermediate value. The peak corresponds to the state where there is a roughly equal fraction of particles in the dilute and dense phases $N_D / N \approx 1/2$. We plot the same for the systems with confinement in Fig. \ref{figReentrant}(c). Notably, the random pinning system also undergoes a transition to a re-entrant fluid. However this transition is delayed relative to the bulk, indicating that the presence of the pins works to stabilize MIPS at high $\pecl$. For the gel system we do not observe the re-entrant behavior within the considered range of $\pecl$.

\section{Conclusion}
\label{sectionConclusion}

Suspensions of Active Brownian Particles show a rich range of dynamical behavior, and our goal was to explore the influence of complex confinement at high densities. To do this, we prepared confining geometries with different static properties: randomly pinned particles from an equilibrium bulk configuration and from a porous gel structure. Both confining geometries are constructed to have the same free volume available to the mobile particles, thus allowing a direct comparison of the effects of the static lengthscale of the confinement on the behavior of active particles.

We first explored the phase behavior at low $\pecl$ revealing how pinning suppressed the crystallization of the fluid at high densities. The relaxation time of the particles is slowed down by the obstacles (more for the random case compared to the gel case) and also it becomes more dynamically heterogeneous, as confirmed by the study of the overlap function and four-point susceptibility.

At intermediate $\pecl$ the bulk system displays MIPS, where the systems forms domains of dense/slow regions and low density/fast regions that nucleate and grow in a similar manner to equilibrium phase separation of two disordered phases. Surprisingly, we find that the obstacles not only do not suppress MIPS formation, but actually act to stabilize it. Random obstacles display a very similar phase-separation pattern as the bulk case, but the domains do not appear homogeneously in the system, but are always formed from the same regions of the sample. The mechanism of MIPS nucleation from random obstacles is still unknown, and should be addressed in future studies. In particular, the effect of system size and changing the state point would be most important to explore. The system size that we have studied here fully demixes to two distinct phases, but the timescale for this for different system sizes, and indeed whether larger systems fully demix would benefit from a detailed finite size analysis. Furthermore, exactly how the pinned particle environment encodes the spatial distribution of the MIPS patterns remains an outstanding challenge. In the case of the gel, the MIPS domains change completely, and forms a 
complex structure where the active and inactive regions occupy different pores of the structure.

Finally we considered how local structure is perturbed by the activity, and revealed that re-entrant MIPS behavior is suppressed (or moved to higher $\pecl$) by the random environments. In particular the gel environment seems to be the most effective in stabilizing the density and activity fluctuations. In this case, we attributed this behavior to the absorption at the rough walls, where the persistent motion of active particles creates localized and highly dense regions, which in turn frees space inside the pores that increase particle transport in the system.

We believe that these results could be important for better understanding transport in biological environments and for guiding future studies of active matter particles in random media. The confining environments that we consider are experimentally realizable \cite{sakai2020}, and indeed combining this with some means to break the symmetry in the future might enable investigations of topotaxis \cite{schakenraad2020}. Other possibilities include the possibility to explore the interplay of the complex environments such as those we have investigated and state functions such as pressure \cite{mallory2014,bialke2015,solon2015} which also might be measured experimentally in colloidal systems \cite{williams2013}.

\section*{Supplementary Material}

Supplementary movie 1: This movie corresponds to the stills in Fig. \ref{figNucleation}. The state point is $\rho=0.87$ and  $\pecl=100$ for the random pinning system. The movie shows the emergence of dense and dilute regions.

\section*{Acknowledgments}
FJM was supported by a studentship provided by the Bristol Centre for Functional Nanomaterials (EPSRC grant EP/L016648/1). 
JR acknowledges support from the European Research Council Grant DLV-759187.
CPR acknowledges support from the European Research Council (ERC Consolidator Grant NANOPRS, project number 617266).

\section*{Appendix}

\subsection*{Overlap}

\begin{figure*}
\centering
\includegraphics[width=\linewidth]{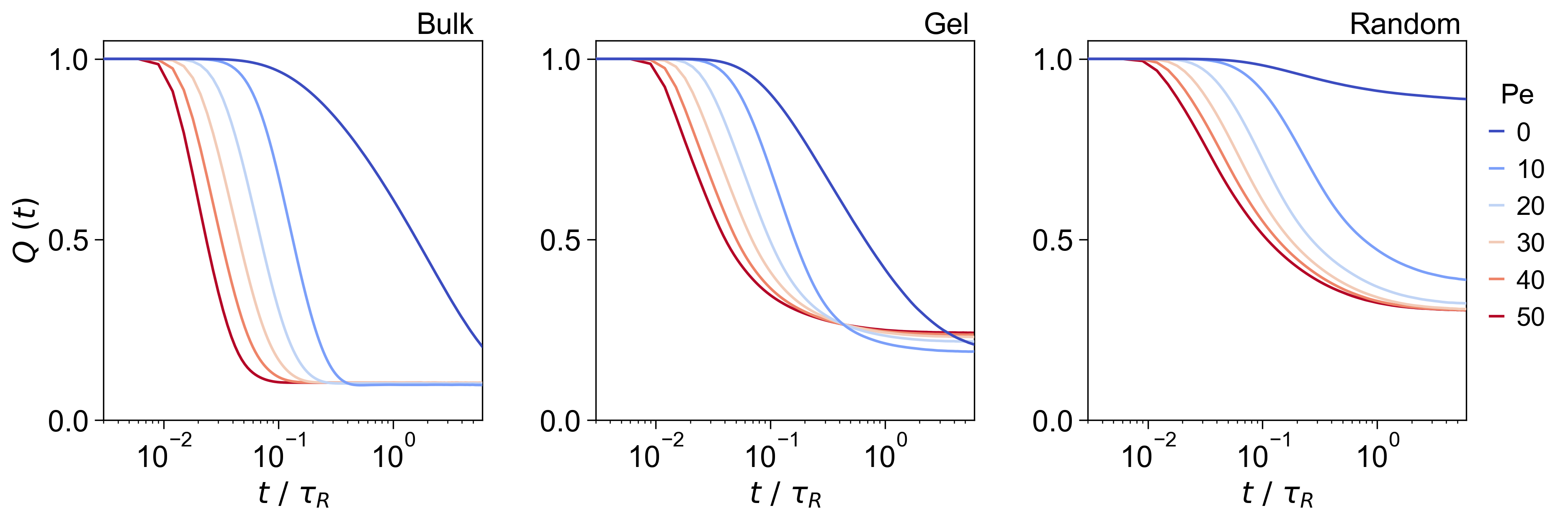}
\caption{The overlap $Q(t)$. Overlap is shown for for particles in the bulk, gel, and random systems at $\rho = 0.87$ for various Pe (see legend).}
\label{figOverlap}
\end{figure*}

The overlap function $Q(t)$ compares a particle configuration with itself at a later time. An important feature of this measure of similarity is that it is not particle specific, it matters not whether it is the same particle occupying that space, only that there is a particle there. With this in mind, it is clear why the bulk system [Fig. \ref{figOverlap}(a)] behaves as it does. In this system, there are no fixed obstacles, and thus there are no points at which a cluster of particles could be anchored. Therefore, even in a system with MIPS present the centre of mass of a dense cluster remains diffusive and thus $Q(t)$ decays to a small value. $Q(t)$ does not decay to $0$ in this case in the bulk due to the relatively high density, as there will always be some degree of overlap.

For active particles in the other two systems, there is clear evidence of the interplay between the self-propulsion and the complex environment. Both systems converge to an overlap value, higher than that of the bulk system. This is a product of the fixed geometry of the obstacles. Although the obstacle particles are not considered, there will be regions in structure that will be more likely to trap active particles, and even though particles are mobile, there is a high probability that there will be some particles in these regions. Interestingly, the gel and the random pins approach their convergent overlaps from different directions as $\pecl$ increases. For the gel system [Fig. \ref{figOverlap}(b)], the overlap value $Q(t)$ increases with $\pecl$ at longer times, while decreasing with an increased rate at shorter times. The increased rate on short timescales is explained by the increased propulsion velocity promoting a quicker re-configuration of the active particle populations. However, this increase in $\pecl$ has the added effect of increasing the likelihood of a particle becoming trapped against the walls of the gel network, this is supported by the increase in the localised fraction  $N_{\textrm{loc}}/N$ with $\pecl$ in Fig. \ref{figLocalised}.

Like in the bulk and the gel network, the overlap of the random system also decays at an increasing rate as $\pecl$ increases at short times (Fig. \ref{figOverlap} c). In the absence of activity, the random system particles have a very slow decay in $Q(t)$. This is a result of the pinned particles dramatically slowing down the dynamics of passive particles, this also seen in the alpha-relaxation time in Fig. \ref{figTauAlpha}, where $\tau_{\alpha}$ was measured to be of the order $10^4$ larger than that of the bulk or gel systems. Unlike in the gel system, $Q(t)$ for the random pinning approaches its convergent value from above. This is likely due to the arrangement of the obstacles in the random pinning system, since the pinned particles are dispersed through the entire space, the active particles have a higher chance of becoming trapped. The random system converges to have higher overlap value than the gel, this is due to the larger fraction of particles becoming localized in the random system compared to the gel at the same activity and density.

\end{document}